\documentstyle[preprint,aps]{revtex}
\begin{document}
\draft
\title{"Centrifugal force: A gedanken experiment"---new surprises}
\author{G. Z. Machabeli}
\address{Department of Theoretical Astrophysics,  Abastumani  Astrophysical
Observatory, Kazbegi str. $N.~2^{a}$, Tbilisi 380060, Republic of
Georgia}
\author{A. D. Rogava}
\address{Department of Theoretical Astrophysics,  Abastumani  Astrophysical
Observatory, Kazbegi str. $N.~2^{a}$, Tbilisi 380060, Republic of Georgia
and Department of Physics, Tbilisi State University,
Chavchavadze ave. 2, 380028 Tbilisi, Republic of Georgia}
\date{\today}
\maketitle

\begin{abstract}

A recently proposed {\it gedanken experiment}  [G.Z. Machabeli and
A.D. Rogava. Phys. Rev. A {\bf 50}, 98 (1994)], exhibiting
surprising behavior, is reexamined. A description of this behavior
in terms of the laboratory inertial frame is presented, avoiding
uncertainties arising due to a definition of a centrifugal force in
relativity. The surprising analogy with the radial geodesic motion in
Schwarzschild geometry is found. The definition of the centrifugal force,
suggested by J.C. Miller and M.A. Abramowicz, is discussed.

\end{abstract}


\section{Introduction}

Recently, in [1], we described a simple
{\it gedanken experiment}, revealing the strange dynamics of rotational
motion in special relativity. The experimental layout consisted of
a straight, long and narrow pipe rotating around an axle normal to its
symmetry axis and  a small bead,  which could move inside the pipe
without friction. The pipe rotated with constant angular velocity
${\omega}=const$ and was assumed to be massles and absolutely rigid.
At $t=0$ the bead was just above the pivot ($r_0=0$) and had an initial
velocity $v_0$. In this particular case the bead, contrary to common
intuitive expectations, appeared to move in a quite unusual way
(see, for details, [1]). The problem was considered in the rotating
non-inertial frame (RNF) of reference of the pipe.

In [1] we have interpreted the surprising behavior of the bead in terms
of reversal in direction of a centrifugal force.
The approach evoked some comments and criticism [2,3].
Miller {\&} Abramowicz in their Comment
[2] pointed at the relationship of this behavior with the relativistic
dependence of the bead mass on its velocity. They recommend to define the
relativistic centrifugal force in the way as forces are defined in general.
Such definition excludes confusing "reversal" of the force, while the
{\it actual deceleration} of the bead is ascribed to the relativistic mass
variation. These interesting comments throw a new light on the subject and
we definitely greet their appearance. However, we would like to reply on the
comments in order to clarify our approach and to point out at some additional
aspects of the problem, which we found out after [1] had appeared in press.

It is, certainly, indisputable that a centrifugal force is a seeming
(or, "apparent" [2]) force, which arises only in non-inertial frames
of reference. However, the main results of [1] were not connected
with the peculiarities of the frame in which the problem was
considered. In this context it seems interesting to reconsider
briefly our experiment in the laboratory inertial frame (LIF),
relative to which the pipe rotates. The purpose is to study the bead dynamics
in this frame, {\it without invoking the centrifugal force conception}. In
the next section such a consideration is presented. The main physically
significant result of [1] (decelarative character of the bead motion) is
verified and obtained again in terms of the LIF.

In the third section we describe an interesting analogy found between
the results of [1] and the radial geodesic motion of the test particle
in Schwarzschild geometry.

In the concluding section of this letter we discuss the definition of
the centrifugal force suggested by Miller {\&} Abramowicz and indicate
at the advantages and disadvantages which, in our opinion, it has.

\section{Bead motion: LIF treatment}

Let us consider the motion of the bead in the LIF, where the spacetime
is just Minkowskian. Owing to the polar symmetry of the experimental
set-up it is convenient for the forthcoming purposes to write the
spatial part of the metric in polar ($g_{{\varphi}{\varphi}}=r^2$,
$g_{rr}=1$) coordinates:
$$
ds^2{\equiv}-d{\tau}^2=-dt^2+r^2d{\varphi}^2+dr^2, \eqno(1)
$$
where $\tau$ is the proper time of the bead. We use units in which $c=1$.

The motion of the bead in LIF is characterised by the three-velocity
${\vec v}$ with the nonzero physical components:
$v{\equiv}v_{\hat r}=dr/dt$ and
$u{\equiv}v_{\hat{\varphi}}=r{\omega}$. Note that $v_{\hat{\varphi}}$
and $v_{\hat{r}}$ are connected with their contravariant and covariant
components as $v_{\hat{\varphi}}=rv^{\varphi}=v_{\varphi}/r$ and
$v_{\hat{r}}=v^{r}=v_{r}$.

The equation of the bead motion in LIF may be written simply as:
$$
{d{\vec p} \over {dt}}={\vec f}, \eqno(2)
$$
where ${\vec p}{\equiv}m_0{\gamma}{\vec v}$ is a three-momentum of the bead,
$m_0$ is its rest mass, and ${\gamma}$--its Lorentz factor as measured in LIF:
$$
{\gamma}={\left(1- {\omega}^2r^2-v^2\right )}^{-1/2}, \eqno(3)
$$
while ${\vec f}$ is a {\it real} three-force acting on the bead
(pipe reaction force). This force has only one, azimuthal, nonzero component:
$f_{\varphi}{\not=}0$. It means that the orthogonal radial component of
the bead three-acceleration $(d{\vec p}/dt)_r$, specified by the left
hand side of (2), is equal to zero. Note that
$(d{\vec p}/dt)_{i}=dp_i/dt+v^kp_{i;k}{\not=}dp_i/dt$
since the spatial part of the metric (1) is curved. In particular, radial
component of the equation of the bead motion leads to
$$
{{d}\over{dt}}(mv)-m{\omega}^2r=0, \eqno(4)
$$
where $m(t){\equiv}m_0{\gamma}$ is the relativistic (inertial) mass of
the bead, which varies with time and measures the beads' {\it variable
resistance} to acceleration. Note that $m(t)$ depends not only on the
radial velocity of the bead $v(t)$, but also on its radial coordinate
$r(t)$. Taking into account (3) we find
$$
{{dm}\over{dt}}=mv{\gamma}^2{\left({\omega}^2 r+
{{dv}\over{dt}}\right)}, \eqno(5)
$$
and (4) yields:
$$
{{d^2r}\over{dt^2}}={{{\omega}^2r}\over{1-{\omega}^2r^2}}
{\left(1-{\omega}^2r^2-2v^2\right)}.
\eqno(6)
$$
This is exactly the same equation, which we get in [1] for the
radial acceleration of the bead as measured in RNF. This equation
also may be written in the following surprisingly elegant form:
$$
{{d^2r}\over{dt^2}}={{(1-{\gamma}^2v^2)}\over{(1+
{\gamma}^2v^2)}}{\omega}^2r. \eqno(7)
$$

This equation distinctly represents peculiarities
of the bead motion: when the motion is nonrelativistic (${\gamma}v{\ll}1$) it
reduces to the usual classic equation for centrifugal acceleration:
$d^2r/dt^2={\omega}^2r$, in the ultrarelativistic limit
($v_0{\to}1$) the sign of the
right hand side is just the opposite: $d^2r/dt^2=-{\omega}^2r$. When
${\gamma}_0v_0=1$ ($v_0={\sqrt 2}/2$) the sign reversal occurs
from the very beginning of the motion.

Another interesting point, which also slipped off
our attention in [1], is that if we introduce new variables:
${\phi}{\equiv}2arccos({\omega}r)$, ${\lambda}{\equiv}{\omega}t$, and
${\Omega}^2{\equiv}1-v_0^2$, we can reduce (6) to the following,
remarkably simple equation:
$$
{{d^2{\phi}}\over{d{\lambda}^2}}+{\Omega}^2sin{\phi}=0. \eqno(8)
$$

This is well-known {\it pendulum equation}, describing nonlinear
oscillations of a free mathematical pendulum. The easiest way for getting
(8) is to write the equation for the radial velocity of the bead [1]:
$$
{dr \over dt} = \sqrt{(1-{\omega}^2r^2)
{\left [1-(1-v_0^2)(1-{\omega}^2r^2) \right ]}},
$$
to rewrite it in above introduced notations as
$d{\phi}/d{\lambda}=-2{\sqrt {1-{\Omega}^2sin^2({\phi}/2)}}$, and to take
one more derivative by $\lambda$. The striking resemblance of
our solutions with mathematical pendulum motion, noticed already in
[1], becomes, now, more clear and appreciable. If we introduce the
concept of an {\it analogous pendulum}, governed by (8), then our
initial conditions ($r_0=0$, $(dr/dt)_{t=0}=v_0$) for this pendulum
are replaced by ${\phi}_0={\pi}$ and
$(d{\phi}/d{\lambda})_{{\lambda}=0}=-2v_0$. This pendulum rotates in
the vertical plane, performing periodic motion with the effective
frequency ${\Omega}$. The time interval,
needed by the bead to reach ${\omega}r=1$ "light cylinder" point [1]
corresponds, now, to the time needed by the analogous pendulum to reach
its stable equilibrium (${\phi}=0$) point.

The LIF treatment has one more advantage: it allows to find out the
pipe reaction force $f{\equiv}f_{\hat{\varphi}}$, which acts on the bead and
forces it to corotate with the rigidly rotating pipe. This force
is explicitly expressed by the azimuthal component of Eq. (3).
The result is:
$$
f={m_0}{\omega}{\left( r{{d{\gamma}}\over{dt}}+2{\gamma}v \right )}
={{2m{\omega}v}\over{1-{\omega}^2r^2}}. \eqno(9)
$$

Note that in RNF this is also the expression for a relativistic
Coriolis (inertial) force acting on the moving bead.

\section{Analogy with the motion in Schwarzschild geometry}

As it appears, this problem is also in a rather unexpected analogy with
the certain kind of geodesic motion in Schwarzschild geometry. In particular,
let us consider a radial geodesic "fall" of a test particle onto a
Schwarzschild black hole with $M$ mass. Let a radial velocity of the particle
at infinity be $V_{\infty}$ and pointed inwards. If one denotes by
$E={\gamma}_{\infty}{\equiv}(1-{V_{\infty}}^2)^{-1/2}>1$ the
specific energy of the particle per its rest mass, then for its radial
velocity relative to the observer at infinity
$V_{\hat{r}}{\equiv}{\sqrt{g_{rr}}}dr/dt$ one gets
$$
{V_{\hat{r}}}^2={E^2 \over 4}-{\left[{E \over 2}-{1 \over E}
{\left(1-{2M \over r}\right)}\right]}^2, \eqno(10)
$$
(see e.g., Mc.Vittie's book [4], where this equation is derived in different
notations). As for the quantity $dV_{\hat{r}}/dt$, called by
McVittie nontenzor radial acceleration of the particle, one gets:
$$
{{dV_{\hat{r}}}\over{dt}}={2M \over {Er^2}}
{\left[{E \over 2}-{1 \over E}{\left(1-{2M \over r}\right)}\right]}
{\sqrt{1-{2M \over r}}}. \eqno(11)
$$

We see that the acceleration of the particle is negative (i.e., the
modulus of the particle infall velocity should be increasing), until the
particle reaches $r_1=4M/(2-{{\gamma}_{infty}}^2)$
radius, where the acceleration changes its sign, and $V_{\hat{r}}$ reaches its
maximum velocity $V_{max}={\gamma}_{\infty}/2$. More specifically, we have
the following kinds of the motion: {\bf(a)}$V_{\infty}{\ll}1$
(${\gamma}_{\infty}{\approx}1$): in this case
the particle begins to move with increasing speed, at $r_1{\approx}=4M$
reaches its maximum value ($V_{max}{\approx}1/2$) and decelerates, afterwards,
down to zero radial velocity, when the particle approaches the black hole
horizon $r{\to}2M$; {\bf (b)}$V_{\infty}={\sqrt 2}/2$
(${\gamma}_{\infty}={\sqrt 2}$): in this "threshold"
case at the initial moment $dV_{\hat{r}}/dt=0$ ($V_{max}=V_{\infty}$,
$r_1 {\to} {\infty}$), and the particle decelerates smoothly down to the
horizon; {\bf(c)}$V_{\infty}>{\sqrt 2}/2$: the acceleration is positive during
the whole course of the motion---the motion is decelerative.

It is easy to notice that this picture impressively resembles (even
quantatively) the situation in our {\it gedanken experiment}. This
likeness is settled due to the remarkable similarity between (10) and
the corresponding equation for $dr/dt$ from [1], which may be written as:
$$
{\left({{dr}\over{dt}}\right)}^2={E^2 \over 4}-{\left[{E \over 2}-{1 \over E}
{\left(1-{\omega}^2r^2\right)}\right]}^2, \eqno(12)
$$

The resemblance is, as it seems likely, a manifestation of some
likeness of the spacetime in the rotating pipe [1]
$$
d{s_p}^2=-(1-{\omega}^2r^2)dt^2+dr^2 \eqno(13)
$$
with the Schwarzschild spacetime along a radial geodesic
(${\theta}=const$, ${\varphi}=const$)
$$
d{s_s}^2=-(1-2M/r)dt^2+(1-2M/r)^{-1}dr^2. \eqno(14)
$$
Comparing, for instance, lapse functions for these two metrics
(${\alpha}_p{\equiv}{\sqrt{-g_{tt}}}={\sqrt{1-{\omega}^2r^2}}$, and
${\alpha}_s{\equiv}{\sqrt{-g_{tt}}}={\sqrt{1-2M/r}}$)
we see that ${\alpha}_p$
and ${\alpha}_s$ become infinite at the light cylinder and horizon,
respectively. However, the likeness is not complete: spatial part
of the pipe metric is flat, while the spatial part of (14) metric is
curved ($g_{rr}{\not=}1$). The latter difference is also essential:
it ensures finitness of the time $t^{*}$ [1], needed by the bead to
reach the light cylinder $r={\omega}^{-1}$. Remember, that an analogous
time to reach the Schwarzschild black hole horizon, as measured by the
distant observer, goes to infinity.

The extensive examination of this analogy needs separate consideration,
which is beyond the scope of this {\it reply} and will be presented elsewhere.

\section{Centrifugal force definition}

It is well known that a generalization of Newton's second law
$$
{\vec F}=d{\vec p}/dt=d(m{\vec v})/dt, \eqno(15)
$$
is the most useful and convenient definition of the (three-) force ${\vec F}$
in special relativity. According to Rindler ([5], p.88) {\it "This definition
has no physical content until other properties of force are specified,
and the suitability of the definition will depend on these other properties."}

For real applied forces, arising in relativistic dynamics, this definition is
physically justified and is proved to be the most appropriate.
But, how one should define {\it inertial} forces in relativity? Miller \&
Abramowicz suggest to use the same general method for these forces too.
We would have nothing against such {\it extrapolation}. However, we
should like to point at one problem arising in the framework of this
approach.

This definition contains a quantity $m(t)=m_0{\gamma}(t)$, which
has, clearly, the meaning of the relativistic mass in the LIF. However,
note that this quantity is used for the definition of
another physical quantity
(centrifugal force), which exists in another---{\it non-inertial}---frame of
reference. Such a procedure (definition of a quantity in one frame,
through some other quantity, defined in  another frame) seems to us
quite unusual for the spirit of special relativity.

Certainly, the variable mass may be defined also in RNF, but
it would not be equal to $m(t)$. The point is that ${\gamma}(t)$, having
in LIF a meaning of Lorentz factor, has not the same meaning in RNF, because
Lorentz factor of a moving particle is not invariant between frames [5].
This circumstance appears mostly obvious in the "$1+1$" formulation of the
same problem. In fact, for the two-dimensional curved metric (13) in the
RNF $V{\equiv}v/{{\alpha}_p}$, and
${\Gamma}=[(1-{\omega}^2r^2)/(1- {\omega}^2r^2-v^2)]^{1/2}$.
Now, it is possible to define relativistic mass in the RNF
as $M(t){\equiv}m_{0}{\Gamma}$ and write, instead of Eq.(15), the
following equation:
$$
F^{*}={{d}\over{dt}}(MV)={{M{\omega}^2r}\over{{{\alpha}_p}}}. \eqno(16)
$$
This definition is already made by means of the true Lorentz factor of the
bead as measured in RNF; like (15) it, also, gives "not reversable"
centrifugal force, but lacks the attractive simplicity of (15).
However, we should always remember that the mass, which the RNF observer
actually {\it measures}, is $M(t)$ and not $m(t)$. It seems, therefore, more
consistent to express the physical quantity existing in a particular frame
(centrifugal force $f_c$, which exists in RNF) through the other physical
quantity $M(t)$ defined and measured in the same frame.

Despite this uncertainity, we should like to note that an importance of the
mass variation effect, noticed by Miller \& Abramowicz, is, indeed, a very
remarkable feature of this problem. As it appears, the capability of mass
variation drastically affects the
dynamics of the motion. A relativistic body is able to "absorb" in
itself an energy, which in nonrelativistic case is expended only on the
increase of its acceleration. But, is it appropriate (and possible) to
describe this secondary dynamical effect as an action of some "negative
self-thrust" force? Saying "secondary" we do not mean its significance but,
rather, its status in the causal order of true physical reasons. If one
introduces such force it, certainly, will be, like the centrifugal force,
an "apparent force". The actual dependence of the bead mass on time is
governed explicitly by the concrete kind of its motion, which is entirely
determined by the outer {\it real} force (pipe reaction force $f$ (12),
in our case) applied to this body.

We would like to finish our reply with modest questions and not with bold
statements. Is it necessary to define seeming, or, "apparent" [2] force
by the rule known for real forces? If, as it was above cited, the
definition is physically aimless until it is not suitable for
describing the properties of a dynamical problem, perhaps it is still
more appropriate to define the centrifugal force in its own way. Maybe,
it is more appropriate to relate the sign reversal of the acceleration
with the corresponding reversal of the properly defined centrifugal
force---the sole inertial force, acting on the moving bead in RNF, which
has the clear nonrelativistic analogy? We hope that this interesting and
intricate problem will attract further more attention of the wide physical
audience.

\section{Acknowledgements}

ADR's research was supported, in part, by International Science Foundation
(ISF) long-term research grant RVO 300.


\begin{references}
\item
G. Z. Machabeli, and A. D. Rogava, Phys.Rev. A {\bf 50}, 98 (1994)
\item
J. C. Miller and M. A. Abramowicz,  SISSA ref. 178/94/A (1994)
\item
F. de Felice, Phys. Rev. A {\bf 52}, 3452 (1995)
\item
G. C. McVittie, {\it General Relativity and Cosmology}, (Chapman and
Hall, London, 1956)
\item
W. Rindler, {\it Essential Relativity}, (Springer-Verlag, New-York
1980)
\end{references}
\end{document}